\documentclass[journal=jpccck,manuscript=article]{achemso}

\usepackage{chemformula} 
\usepackage[T1]{fontenc} 
\usepackage{ulem}

\author{Abhishek Kumar}
\affiliation{NEST, Istituto Nanoscienze--CNR and Scuola Normale Superiore, Piazza San Silvestro 12, 56127 Pisa, Italy}

\author{Francesca Telesio}
\affiliation{NEST, Istituto Nanoscienze--CNR and Scuola Normale Superiore, Piazza San Silvestro 12, 56127 Pisa, Italy}

\author{Deborah Prezzi}
\affiliation{S3, Istituto Nanoscienze-CNR, Via Campi 213/A, 41125 Modena, Italy}
\email{deborah.prezzi@nano.cnr.it}

\author{Claudia Cardoso}
\affiliation{S3, Istituto Nanoscienze-CNR, Via Campi 213/A, 41125 Modena, Italy}

\author{Alessandra Catellani}
\affiliation{S3, Istituto Nanoscienze-CNR, Via Campi 213/A, 41125 Modena, Italy}

\author{Stiven Forti}
\affiliation{Center for Nanotechnology Innovation @ NEST, Istituto Italiano di Tecnologia, Piazza San Silvestro 12, 56127 Pisa, Italy}

\author{Camilla Coletti}
\affiliation{Center for Nanotechnology Innovation @ NEST, Istituto Italiano di Tecnologia, Piazza San Silvestro 12, 56127 Pisa, Italy}

\author{Manuel Serrano--Ruiz}
\affiliation{CNR-ICCOM, Via Madonna del Piano 10, 50019 Sesto Fiorentino, Italy}

\author{Maurizio Peruzzini}
\affiliation{CNR-ICCOM, Via Madonna del Piano 10, 50019 Sesto Fiorentino, Italy}

\author{Fabio Beltram}
\affiliation{NEST, Istituto Nanoscienze--CNR and Scuola Normale Superiore, Piazza San Silvestro 12, 56127 Pisa, Italy}

\author{Stefan Heun}
\affiliation{NEST, Istituto Nanoscienze--CNR and Scuola Normale Superiore, Piazza San Silvestro 12, 56127 Pisa, Italy}
\email{stefan.heun@nano.cnr.it}

\title{Black Phosphorus n-type doping by Cu: \\ a microscopic surface investigation}


\begin{document}

\begin{tocentry}
\centering
\includegraphics{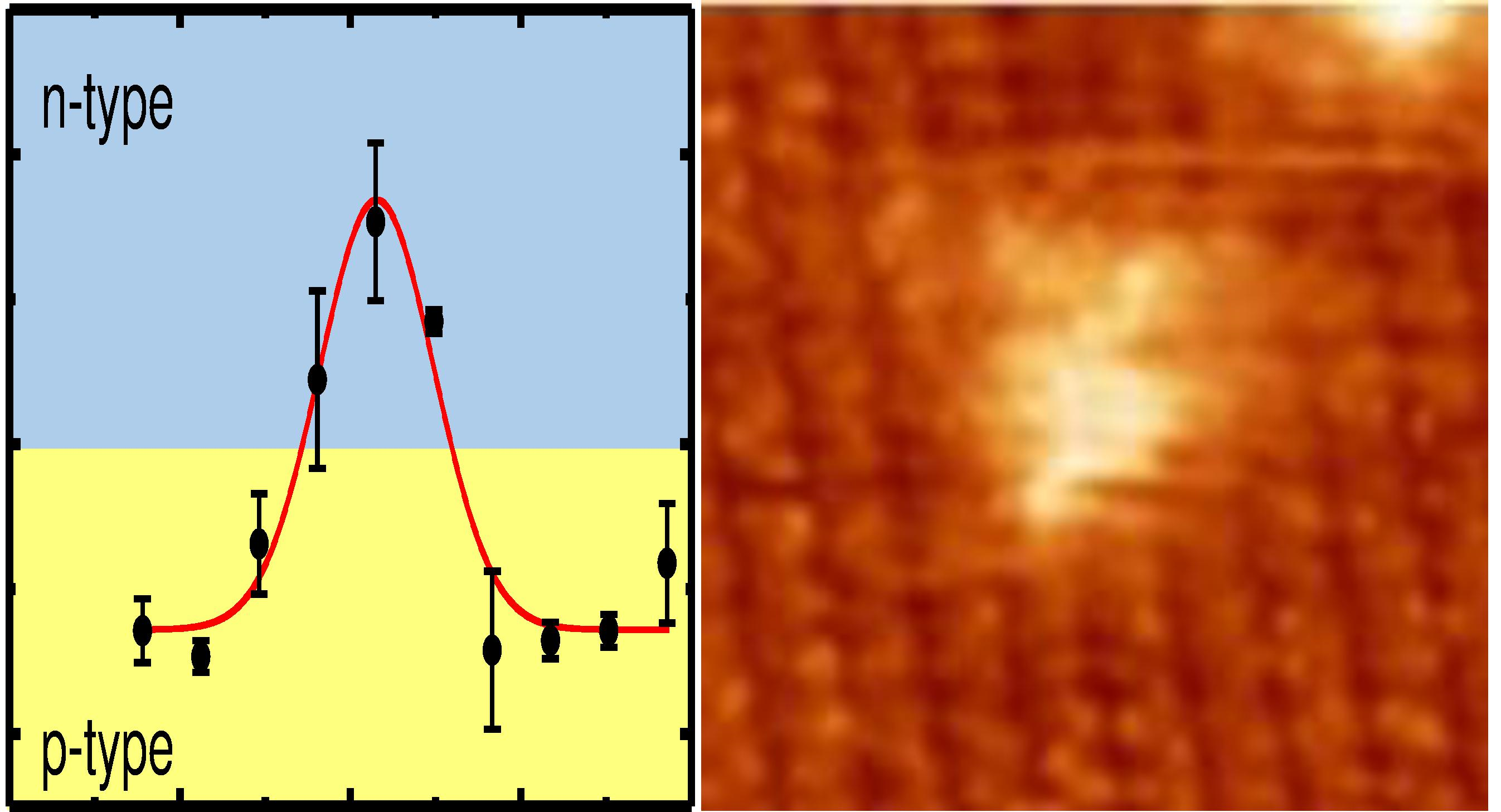}
\end{tocentry}

\begin{abstract}
We study surface charge transfer doping of exfoliated black phosphorus (bP) flakes by copper using scanning tunneling microscopy (STM) and spectroscopy (STS) at room temperature. The tunneling spectra reveal a gap in correspondence of Cu islands, which is attributed to Coulomb blockade phenomena. Moreover, using line spectroscopic measurements across small copper islands, we exploit the potential of the local investigation, showing that the n-type doping effect of copper on bP is short-ranged. These experimental results are substantiated by first-principles simulations, which quantify the role of cluster size for an effective n-type doping of bP and explain the Coulomb blockade by an electronic decoupling of the topmost bP layer from the underlying layers driven by the copper cluster. Our results provide novel understanding --difficult to retrieve by transport measurements-- of the doping of bP by copper, which appears promising for the implementation of ultra-sharp p-n junctions in bP.
\end{abstract}

\section{Introduction}

Black phosphorus (bP) is a semiconductor with a direct band gap that ranges from $\sim 0.3$~eV (bulk) to $\sim 2.0$~eV (monolayer) depending on layer thickness.\cite{Das2014,Tran2014,Roldan2017} Among the van der Waals elemental materials,\cite{Ling2015,Peruzzini2019} bP in its few-layer form attracted great interest since its first exfoliation \cite{Gomez2014,Liu2014} because of the modulation of the direct band gap,\cite{Ling2015,Churchill2014} the in-plane anisotropy,\cite{Xia2014c,Lee2015a,Telesio2019a} and its high charge-carrier mobility, up to 5,200~cm$^{2}$/(V$\cdot$ s) at room temperature,\cite{Long2016} appealing for possible device applications.\cite{Huang2019}

The presence of phosphorus vacancies, as reported in a recent scanning tunneling microscopy (STM) study,\cite{Kiraly2017} makes bP an intrinsically p-doped  material.\cite{Morita1986,Narita1983} N-type doping of bP has been investigated in the light of possible applications.\cite{Hu2020} Several strategies have been pursued, from the reversible doping by field effect in transistors, where an ambipolar behaviour was obtained,\cite{Das2014,Perello2015,FieldDopingbP2017} to exfoliation of bulk crystals obtained by substitutional doping with Se \cite{Xu2016Se} or Te.\cite{Yang2016Te} A more promising strategy for few-layer flakes could be surface charge transfer doping, which has been implemented on several two-dimensional materials such as graphene \cite{Chen2007,Starke2012a}  and transition metal dichalcogenides.\cite{Zhang2018,Xu2017} For bP, few studies report n-type doping by surface charge transfer,\cite{MoO32015,Kim2015,Koenig2016,Yu2016,Cho2017a,Gao2018a,Kiraly2019,Lin2019,Ou2019a} mostly for the case of alkali metals.\cite{Kim2015,Kiraly2019,Gao2018a} Electrical transport measurements show n-type behavior when bP is doped by the deposition of copper.\cite{Koenig2016,Lin2019} However, no local spectroscopic investigation has been performed so far to understand the effect of Cu on the electronic properties of bP at the atomic level.  

In this work, we performed an investigation of surface charge transfer doping by copper deposition on exfoliated bP flakes, combining STM and scanning tunneling spectroscopy (STS) measurements with first-principles calculations based on density functional theory (DFT). STS shows a gap in the spectra measured on Cu islands that we attribute to Coulomb blockade. The data also suggest that Cu induces an n-type doping in bP. Line-spectroscopic measurements across copper islands further show that the copper doping effect on bP is very short-ranged. Theoretical simulations reveal that Cu clusters efficiently decouple the topmost bP layer via charge localization, which explains both the Coulomb blockade as well as the local doping effect observed in experiments.  

\section{Results and Discussion}

\begin{figure}
   \centering
   \includegraphics[width=0.5\columnwidth]{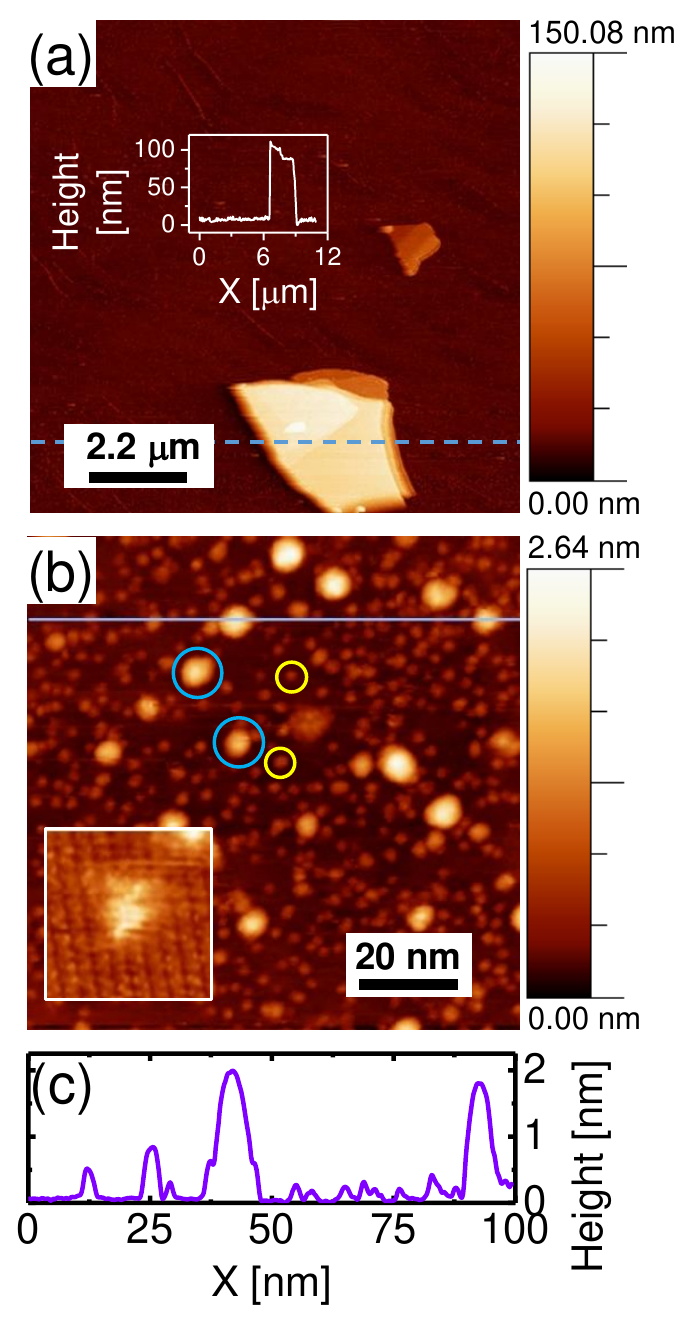}
   \caption{\label{Fig1} (a) 11~$\mu$m~$\times$~11~$\mu$m STM image showing two bP flakes on a graphene substrate after 20~minutes of copper deposition. Scan parameters: ($-3.0$~V, $0.33$~nA). Inset: height profile across the larger flake along the dashed blue line. {The flake has two plateaus; the higher region (top part of the flake) is 93 nm high, the lower (bottom part) 80 nm. All measurements presented in the following panels were performed on the lower plateau.} (b) 100~nm~$\times$~100~nm STM image on the larger flake, showing copper islands on bP. Some of the smaller and larger copper islands are identified by yellow and blue circles, respectively. Inset: 5~nm~$\times$~5~nm STM image obtained upon further zoom-in, showing a copper island on top of the atomically resolved bP surface. Scan parameters for both images: ($-1.0$~V, $0.33$~nA). (c) Height profile along the violet line in (b).}
\end{figure}

Figure~\ref{Fig1} reports an STM analysis of copper deposited on bP flakes supported by a graphene-on-SiC substrate, which acts as a conductive ground electrode.\cite{Abhishek2018} A comparison of bP surfaces before and after copper deposition is shown in the Supporting Information (Figure~S2), where the pristine bP surface appears clean and flat at this magnification, with some defects related to intrinsic phosphorus vacancies, consistent with previous reports.\cite{Kiraly2017,Abhishek2018,Riffle2018} Figure~\ref{Fig1}a shows two representative bP flakes on which copper was deposited for 20 minutes. A zoom-in on the larger flake (Figure~\ref{Fig1}b) shows bright features on the surface that appear only after copper deposition and are therefore identified as copper islands, {indicative of a Volmer-Weber growth mode}. Copper islands of two different sizes can be seen, indicated by blue (larger islands, diameter $(7.84 \pm 1.08)$~nm, height $(1.91 \pm 0.31)$~nm) and yellow circles (smaller islands, diameter $(3.75 \pm 0.51)$~nm, height $(0.63 \pm 0.11)$~nm). From a statistical analysis of several different spots on the flake, we compute a coverage of $(0.80 \pm 0.11)$~ML of copper on the bP surface \bibnote{To quantify copper coverage, we calculated the volume of copper islands in STM images by flooding analysis and normalized it by the amount of copper required for a uniform monolayer (ML) coverage. We define 1 ML as $1.767 \times 10^{15}$ atoms/cm$^{2}$. To get this number, we used 8.96~g/cm$^{3}$ as the copper density, $1.055 \times 10^{-22}$ g as copper atomic mass, and $2.08$ {\AA} as the spacing between (111) planes in the Cu fcc lattice \cite{Davey1925}}. We underline that these islands are rather different from the chiral Cu structures that were recently reported and were obtained upon heating of Cu nanoparticles on bP flakes at 300 $^\circ$C.\cite{Nerl+19afm}

A further zoom-in on the flake is shown in the inset of Figure~\ref{Fig1}b. The surface shows the zigzag pattern characteristic of the [100] direction of bP \cite{Abhishek2018} and a copper island on top of it. {This observation  allows to exclude an intercalation of the Cu here since in that case the typical lattice structure of bP should be visible also on top of the intercalated copper, as observed in similar systems such as Lithium-intercalated graphene \cite{Fiori2017}.} The fact that the edges of the copper island appear fuzzy when imaged at high resolution is a sign of thermally-activated motion of copper atoms at room temperature.\cite{Sobotik2018} Despite this motion, however, the copper islands are stable, since their relative positions remain unchanged in multiple scans of the same location (Figure~S3). This excludes any tip-induced effects and is in contrast to other reports, where motion or modification of islands on the substrate was induced by the tip of an STM.\cite{Murata2019,Buech2018}

\begin{figure*}[t]
   \includegraphics[width=\textwidth]{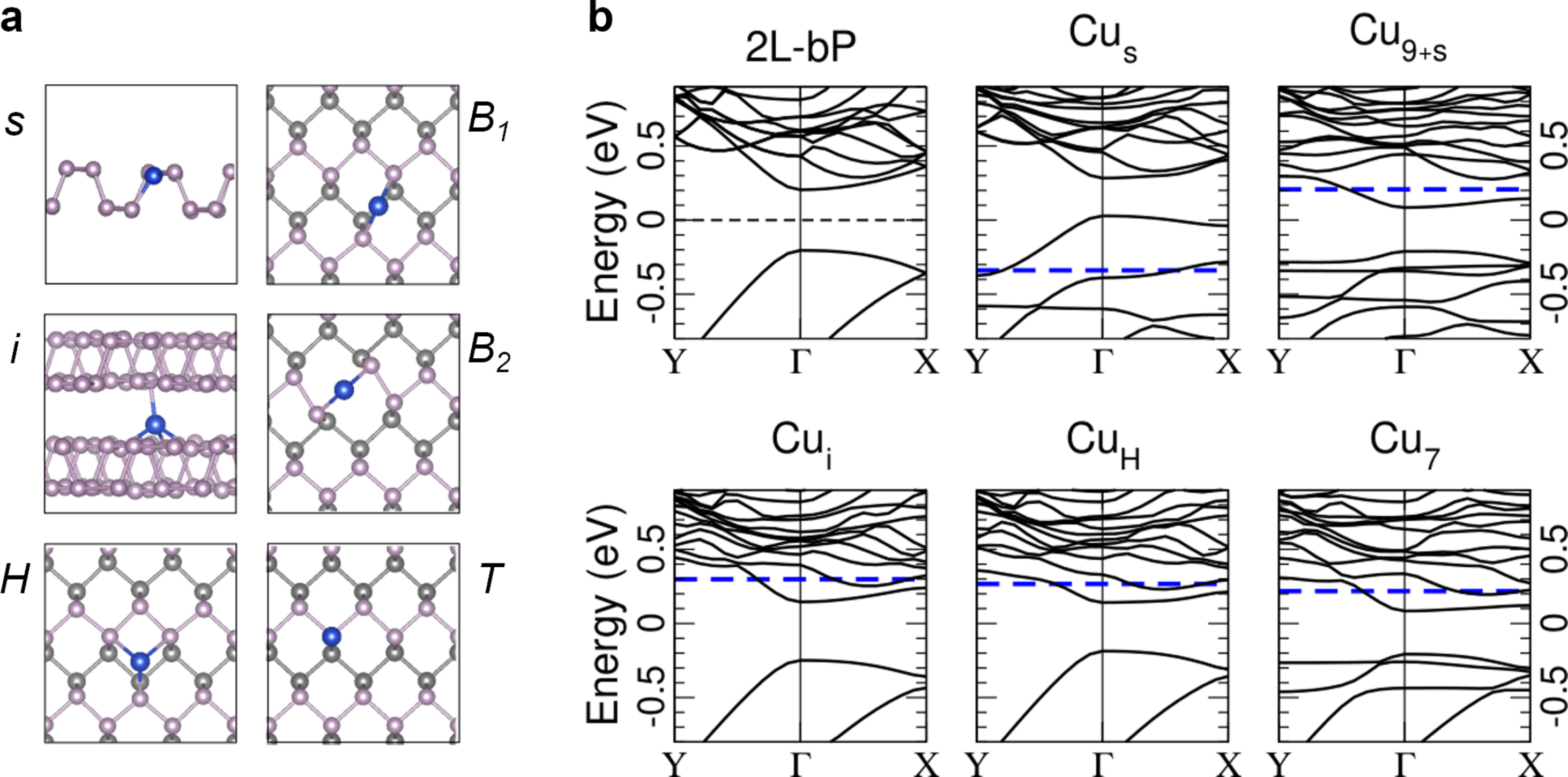}
   \caption{\label{fig:struct-bands} (a) Ball-and-stick models for single Cu impurities (blue spheres) in bP. The interstitial, $i$, and substitutional, $s$, sites are shown as side views. For adsorbed Cu atoms, several sites were considered, that is hollow ($H$), top ($T$) and two different bridge positions ($B_1$ and $B_2$), shown as top views. For the top views, the bottom phosphorous layer is represented in grey, in order to make the figures clearer. (b) Band structure of the three lowest energy Cu impurities and two representative Cu clusters. The band structure of pristine 2L-bP is reported for comparison, the midgap of which is set as the zero energy reference for all band structures. The Fermi level of the different Cu configurations is highlighted by dashed blue lines. {The  high-symmetry  points  X  and  Y  indicate  the zone-center along the zigzag and armchair direction, respectively.}}
\end{figure*}

This structural microscopic investigation by STM is coupled to DFT-based ab-initio simulations addressing the formation energies of the main Cu point defects and adatoms on few-layer bP slabs, together with adsorbed clusters of increasing size. Figure \ref{fig:struct-bands}a illustrates the single Cu impurities investigated here: substitutional (Cu$_s$), interstitial (Cu$_i$), and adsorbed copper in four different adsorption sites, i.e.~hollow (Cu$_H$), top (Cu$_T$) and two bridge sites (Cu$_{B_1}$ and Cu$_{B_2}$, bridging atoms on adjacent zigzag rows and adjacent atoms on the same zigzag row, respectively). According to the values reported in Table \ref{tab:formation_energy}, Cu$_s$ saturating P vacancies is the most stable point defect, followed by Cu$_i$, while single Cu adatoms are preferentially adsorbed in a hollow position (Cu$_H$).  

\begin{table}
\centering
   \caption{\label{tab:formation_energy} Formation energy ($\Delta$E) of Cu single impurities (left) and Cu clusters (right) on bP at $T=0$~K. For Cu adatoms, several sites were considered, that is hollow ($H$), top ($T$) and two different bridge positions ($B_1$ and $B_2$). The formation energy of adsorbed Cu clusters of increasing size is reported per Cu atom, while the energy gain with respect to the single impurities is per Cu adsorbed  atom. Clusters are labeled as Cu$_{n+s}$ ($n$ Cu atoms added to the Cu$_s$ impurity) and Cu$_{n}$ ($n$ adsorbed Cu atoms including the initial Cu$_H$).}
   \begin{tabular}{lc|lcc}
      \hline
      &\\[-5pt]
      site  &  $\Delta$E (eV) & cluster & $\Delta$E/N$_{\textrm{Cu}}$ (eV) & $\delta$E/N$_{\textrm{Cu}_{ads}}$ (eV)\\[5pt]
      \hline\hline
      &\\[-5pt]
      $s$     & $-4.47$   & Cu$_{1+s}$ & $-4.01$ & $-0.87$\\
      $i$     & $-3.46$   & Cu$_{3+s}$ & $-3.56$ & $-0.58$\\
      $H$     & $-2.68$   & Cu$_{7+s}$ & $-3.36$ & $-0.51$\\
      $B_1$   & $-2.55$   & Cu$_{9+s}$ & $-3.42$ & $-0.62$\\
      $B_2$   & $-1.61$   & Cu$_3$     & $-2.85$ & $-0.08$\\
      $T$     & $-1.51$   & Cu$_7$     & $-3.03$ & $-0.35$\\[5pt]
      \hline
   \end{tabular}
\end{table}

The formation of clusters on the bP surface is also investigated, as shown in Table \ref{tab:formation_energy}, starting from the most stable single Cu impurities decorating the surface, i.e.~Cu$_s$ and Cu$_H$. In particular, we considered two series of clusters as obtained by using either a surface substitutional Cu$_s$ or a single adsorbed Cu$_H$ atom as nucleus and adding further Cu atoms to build clusters of increasing size. The two series are hereafter labeled as Cu$_{n+s}$ ($n$ Cu atoms added to the Cu$_s$ impurity) and Cu$_{n}$ ($n$ adsorbed Cu atoms including the initial Cu$_H$). In addition to the formation energy per Cu atom, Table \ref{tab:formation_energy} reports also the energy gain for the clusters with respect to the isolated impurities per adsorbed Cu atom (last column): this value indicates that cluster formation is always favored, in overall agreement with experimental observations, which do not evidence single Cu atoms on the surface at room temperature. More specifically, we find that cluster nucleation around Cu$_s$ sites is the (thermodynamically) most favorable process.  

For a better understanding of the electronic properties of Cu on bP, we report in Figure \ref{fig:struct-bands}b the computed DFT band structure for the three lowest-energy isolated Cu impurities discussed above, i.e.~Cu$_s$, Cu$_i$ and Cu$_H$, together with the band structure of selected adsorbed clusters. The band structure of pristine 2-layer (2L-) bP is shown as a reference. Starting with the single Cu impurities, we notice that Cu$_i$ and Cu$_H$ behave as n-type dopants whereas substitutional copper leads to a p-doping of bP. While for the n-doped systems (Cu$_i$ and Cu$_H$) the valence band is very similar to the pristine one, the p-doped system (Cu$_s$) has a markedly different valence-band structure. In all cases, the band gap is either reduced (by up to 150 meV for Cu$_s$) or almost unchanged (Cu$_i$) \bibnote{We note that structural optimization gives rise to a sliding of the layers in presence of interstitial Cu and a subsequent gap opening due to a partial decoupling of the layers. This behaviour is not expected to show up in real samples. We have thus chosen to mimic real samples by inhibiting layer sliding. In the latter case we find no significant gap opening} in the presence of Cu atoms, as previously observed for other TM adatoms.\cite{babar2016,Hu2015b}

\begin{figure*}[t]
    \includegraphics[width=\textwidth]{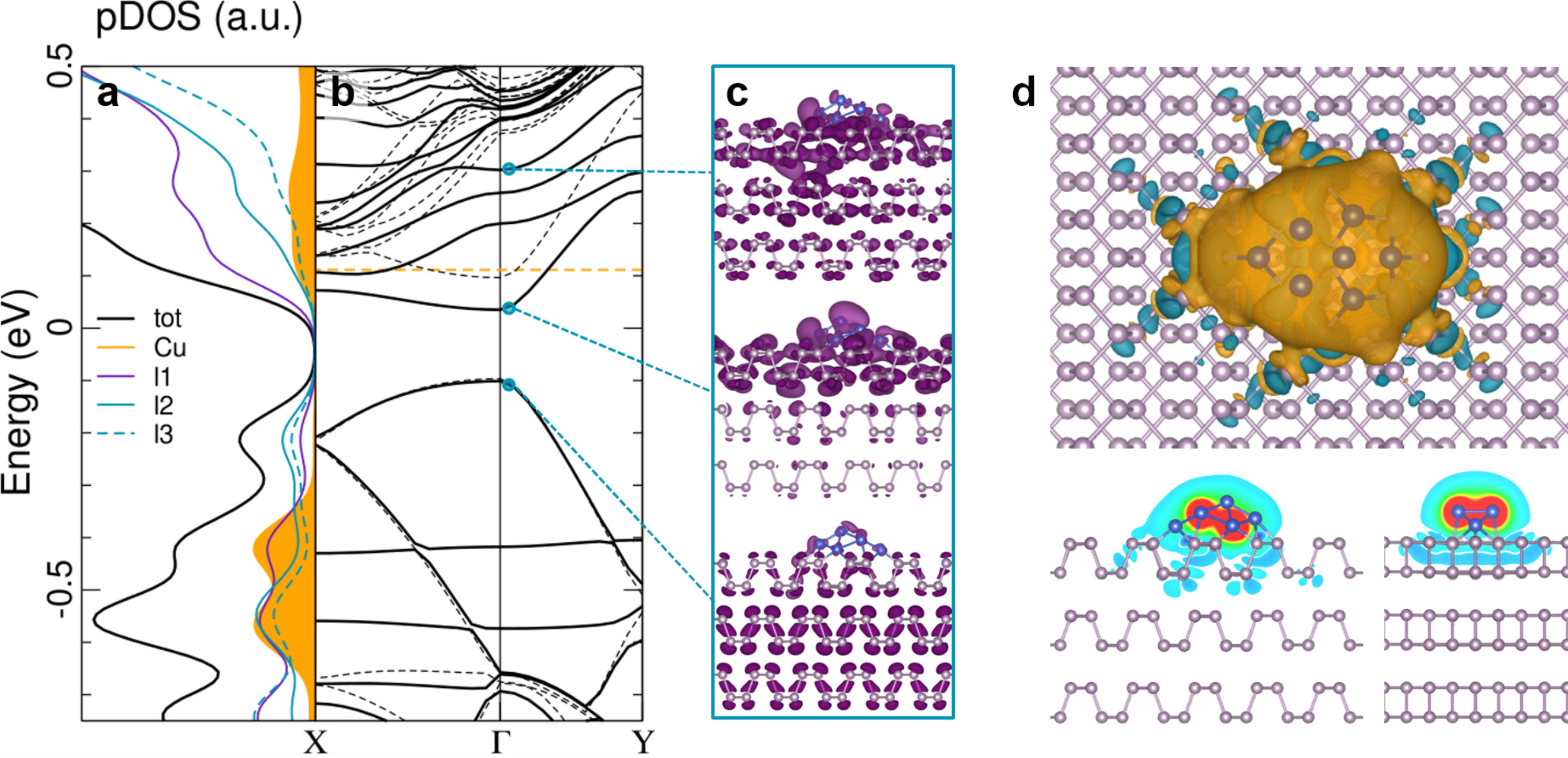}
    \caption{\label{fig:wavefuntions_7} Electronic properties of adsorbed Cu$_7$ on 3L-bP. (a) The total density of states (black line) is split into the contribution of the Cu cluster (orange curve) and bP first (l1), second (l2) and third layers (l3) by using atomic projections. (b) The band structure in presence of the adsorbed Cu cluster (solid black line) is compared to that of pristine 3L-bP (dashed line). The Fermi level of the doped system is also reported (orange dashed line), with the midgap of the pristine 3L-bP set as the zero energy reference. (c) A few relevant Kohn-Sham squared wave functions are displayed, with purple (blue) balls denoting P (Cu) atoms. The isosurface amplitude is $7\times 10^{-5}$ a.u. (d) Top-view of the charge density difference with an isosurface amplitude of $3\times 10^{-4}$ a.u. (top panel, orange (blue) is the positive (negative) isosurface), together with two sections of the isosurface along armchair and zigzag directions (bottom panel). For the sections, we use a RGB color scale with red and blue corresponding to 0.03 and -0.008, respectively.}
\end{figure*}

Considering now the clusters, the doping effect for Cu$_7$, which is adsorbed on the pristine surface, is very similar to that of a single Cu adatom (Cu$_H$), i.e.~n-type, and shows only some additional nondispersive Cu states in the region close to the band gap. In the case of clusters building on a substitutional Cu$_s$ seed (Cu$_{n+s}$), the n-type doping effect of the cluster tends to overcome the p-type doping character of the Cu$_s$ seed for sufficiently large clusters, resulting in a band structure similar to that of the adsorbed clusters without seed, as can be seen by comparing Cu$_{9+s}$ and Cu$_7$. This points to the existence of a critical size at which all Cu clusters would behave the same, i.e.~n-doping and band gap reduction, irrespective of the presence of a substitutional seed.  

We further investigate the band gap modification and doping induced by copper clusters by focusing on Cu$_7$ and consider a larger supercell (5 $\times$ 5) and a thicker bP slab (3-layer, 3L) to minimize the effect due to periodic replicas. A similar analysis for the case of Cu$_{7+s}$ is reported in the Supporting Information (Figure~S4). Figure~\ref{fig:wavefuntions_7}a shows the projected density of states (pDOS) of Cu$_7$ on 3L-bP separated into the contribution of Cu atomic orbitals (orange curve) and the atomic orbitals of each bP layer (violet and light blue). Panel b shows the band structure computed for the Cu$_7$/3L-bP case, as compared to the band structure of pristine 3L-bP (black dashed lines). While the band structure highlights the same qualitative features described above for Cu$_7$/2L-bP (see Figure \ref{fig:struct-bands}), the pDOS shows that the copper affects mainly the topmost bP layer, which presents Cu-derived peaks at about $-0.4$ and $+0.15$~eV (l1, violet line). At these energies, the contributions from the bottom bP layers (l2 and l3, light blue) are instead significantly smaller. This suggests that the Cu levels around the band gap region hybridize mainly with states localized in the topmost bP layer and are essentially decoupled from the layers underneath. In contrast, deeper Cu-originating states hybridize with states from all three bP layers.  

This is confirmed by the plot of the Kohn-Sham orbitals for selected states (Figure~\ref{fig:wavefuntions_7}c). Indeed, the orbitals at the bottom of the conduction band are spatially localized at the topmost bP layer. Only at higher energies, the wavefunctions are distributed over the entire system (i.e.~similar to those of pristine bP). Deeper insight into the charge-density redistribution upon Cu doping can be obtained by plotting the charge-density difference for the Cu$_7$/3L-bP system, obtained by subtracting the charge computed for the pristine substrate from the total charge of the doped system (see Figure~\ref{fig:wavefuntions_7}d). From both the top view of the isosurface plot (top panel) and the isosurface sections along the zigzag and armchair directions (bottom panels), we note that: (i) the Cu charge transferred to bP remains mostly localized on the topmost layer; (ii) the charge shows a strong localization in the bP plane around the cluster, {with an abrupt decay within a few lattice units.} In addition, the charge rearrangement in the stacking direction reveals the origin of the spatial localization of the frontier orbitals described above as a result of the local field generated at the interface, reminiscent of what is found in presence of an applied external electric field.\cite{liu+15nl,Dolui2015} This finds a correspondence in the potential energy difference (or workfunction difference) between the top (i.e.~doped side) and bottom of the Cu-decorated bP slab, which is positive for the n-type doped Cu$_7$/3L-bP system and amounts to 98 meV.  

\begin{figure*}[!t]
   \includegraphics[width=\textwidth]{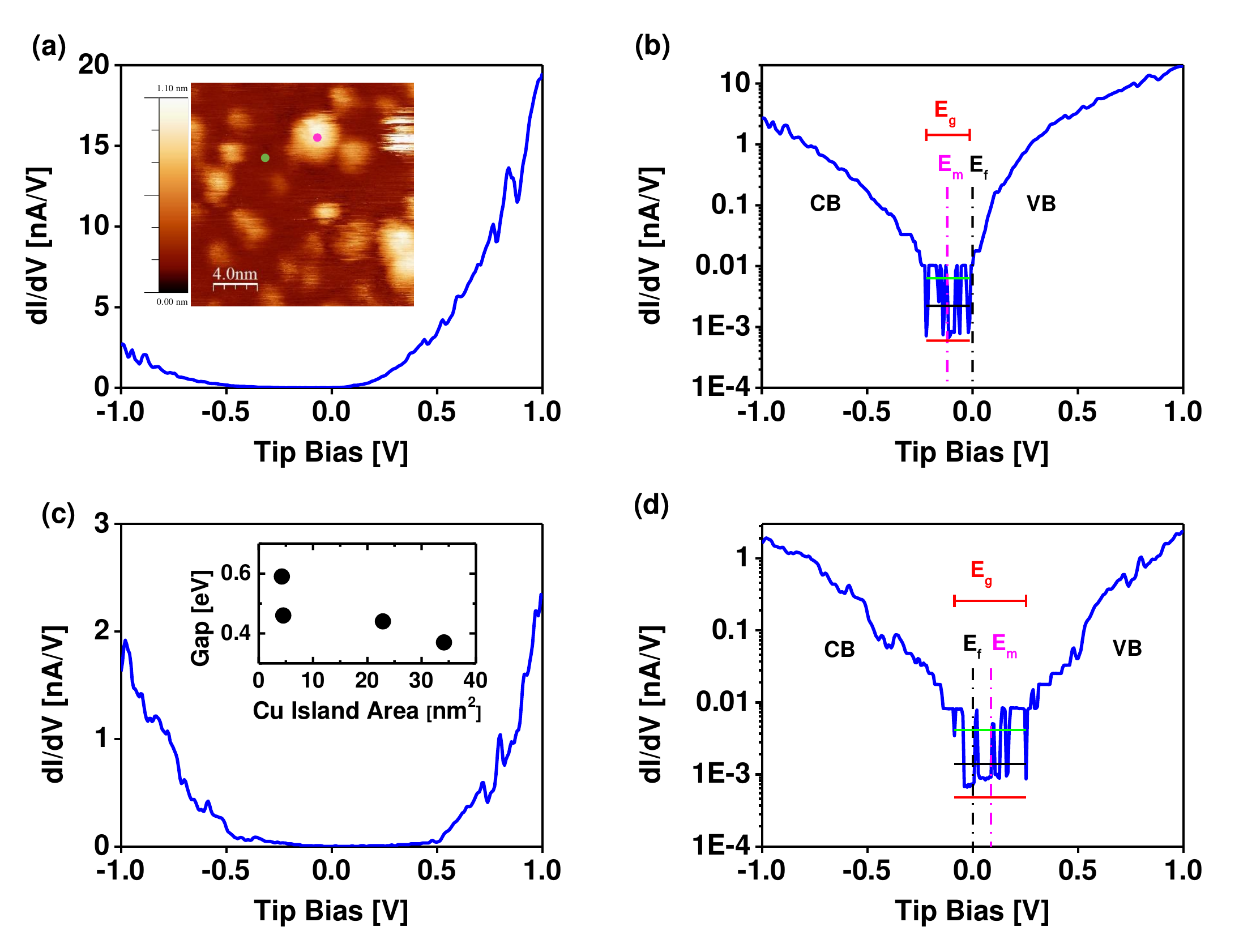}
   \caption{\label{Fig2} Differential conductance (dI/dV) spectra (a) for pristine bP and (c) for Cu on bP. The green and the pink spots in the STM image shown as inset to (a) indicate where the spectra in (a) and (c), respectively, were recorded. {The same spectra as in (a) and (c)} are plotted on a logarithmic scale in (b) and (d) for pristine bP and Cu on bP, respectively. The inset to (c) shows how the gap measured on Cu islands changes with island area. Scan parameters of the STM image in (a) and STS setpoint for all spectra (on bP and Cu islands): ($-1.0$~V, $0.33$~nA). The bias is applied to the tip.}
\end{figure*}

Taking advantage of the DFT analysis, we next move to local spectroscopy by STS, both on flat bP surface areas and on copper islands. Typical differential-conductance curves are shown in Figure~\ref{Fig2}, where both linear and logarithmic plots are displayed for an analysis of spectral shape and identification of the gap region, respectively. The  shape of the spectra recorded off the copper islands (Figure~\ref{Fig2}a), with a larger amplitude at positive tip bias, very much resembles that of pristine bP.\cite{Zhang2009,Kiraly2017} On the contrary, the spectra measured on copper islands (Figure~\ref{Fig2}c) display similar spectral amplitude at $\pm 1$ V and thus a more symmetric shape. We can exclude that the shape of the spectra is affected by thermal drift of the STM tip in z-direction, since we have checked that for all spectra the forward and backward sweeps were coinciding (as shown in the Supporting Information).

{From more than 40 spectra measured in several spots on the flake,} we obtain a gap value of $(0.25 \pm 0.10)$~eV for the flat bP areas, consistent with the reported band gap value of pristine bP.\cite{Morita1986,Zhang2009,Kiraly2017,Riffle2018} As shown in Figure~\ref{Fig2}b, the position of the Fermi level (at zero bias, vertical black dot-dash line) is lying close to the valence-band edge, which implies p-type doping, consistent with the literature for pristine bP.\cite{Kiraly2017} Thus, the spectra measured on the flat bP surface away from Cu islands can be attributed to pristine bP. The same analysis for spectra recorded on copper islands gives a gap value of $(0.46 \pm 0.20)$~eV. {To determine the gap values, we followed the procedure described in Ref. \cite{Kiraly2017}.}

Even though the increase in the band gap and the symmetry of the differential-con\-duc\-tance curves in the presence of the Cu islands are in apparent contradiction with the DFT results described above, the results can be consistently understood if we invoke a Coulomb blockade for the Cu islands. Similar effects have already been observed in the single-electron tunneling spectra of ultrasmall metal islands at room temperature.\cite{Schoenenberger1992a,Schoenenberger1992b,Dorogi1995} Here, DFT calculations show that close to the Fermi level, the localized frontier states do not hybridize with the lower bP layers, thus creating a tunnel barrier between the Cu islands and the bP bulk. In this situation, the electron charge on the Cu islands can be quantized if two conditions are met:\cite{Schoenenberger1992a,Oncel2005} (i) The capacitance $C$ of the islands is sufficiently small such that the charging energy $e^2/2C$ exceeds the thermal energy $k_B T$. Here, applying a parallel plate capacitor model,\cite{Hong2013} we estimate $C \approx 10^{-18}$~F, which leads to $e^2/2C \approx 100$~meV, larger than $k_B T$ ($\approx 25$~meV at room temperature). (ii) The tunneling resistance $R_T$ between STM tip and Cu island is much larger than the resistance quantum $R_K = h/e^2 \approx 25.8$~k$\Omega$. From Figure~\ref{Fig2}d, we get $R_T > 10$~G$\Omega$ for $V_{tip} < \pm 0.5$~V. Being in a situation in which these two conditions are met, it is reasonable to conclude that the experimentally observed gap {might be} due to Coulomb blockade. Consistently, a  Coulomb gap of 0.45~eV ($=e/C$)\cite{Hong2013} corresponds to a capacitance $C = 4 \times 10^{-19}$~F, in good agreement with the estimate based on the parallel plate capacitor model. A further evidence of Coulomb blockade effect is given by the gap dependence with the island dimension. As shown in the inset to Figure~\ref{Fig2}c, the experimentally observed gap decreases with increasing island area, as expected for Coulomb blockade,\cite{Brun2012} approaching the value of the bP band gap for the largest island measured.  

The STS data in Figure~\ref{Fig2}d, which was measured on a Cu island, shows that the position of the Fermi level does not coincide with the midgap position but is shifted towards the conduction band edge. According to the orthodox theory of tunneling through a double junction, which applies here,\cite{Brun2012} the asymmetric gap observed in Figure~\ref{Fig2}d is due to the fractional residual charge $Q_0$ on the Cu island.\cite{Hanna1991} According to Hanna and Tinkham,\cite{Hanna1991} $Q_0$ originates from the difference in work function, or the contact potential, across the junctions. $Q_0$ is then obtained from
\begin{equation}
    Q_0 = \frac{1}{e} \left[ C_1 \Delta \Phi_1 - C_2 \Delta \Phi_2  \right]
\end{equation}
with $C_i$ the capacitance of junction $i$ and $\Delta \Phi_i$ the difference in work function across junction $i$; $i = 1$ indicates the tip-island junction and $i = 2$ the island-substrate junction. Here, we observe a positive $Q_0 > 0$, thus a positive value of the contact potential, which indicates an n-type doping of the bP, in agreement with our DFT-calculations and literature.\cite{Koenig2016,Lin2019}

\begin{figure}
   \centering
   \includegraphics[width=0.5\columnwidth]{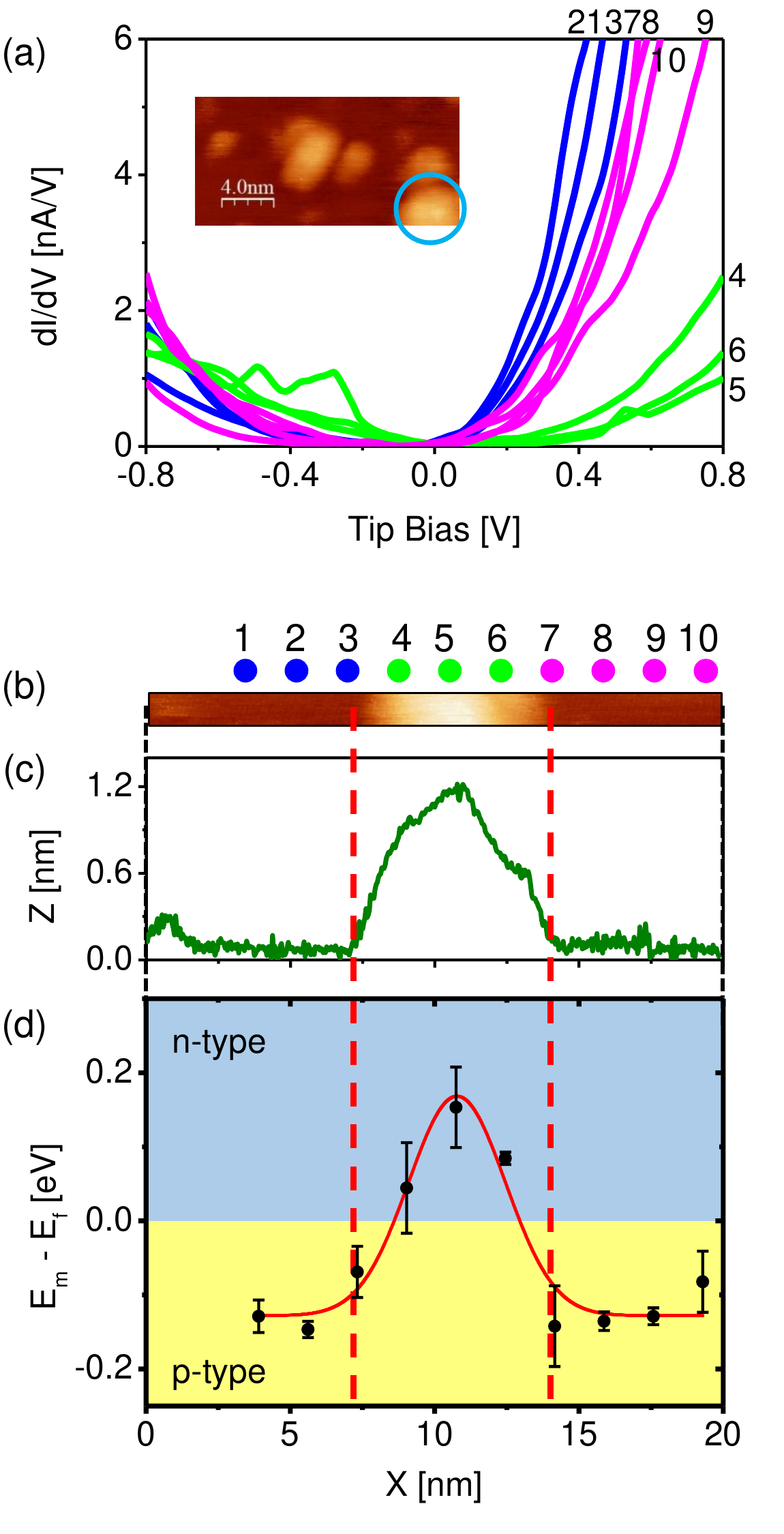}
   \caption{\label{Fig3} Line STS across a single copper island. (a) Average spectra recorded along a line across a copper island, numbered corresponding to lateral spots shown in (b). The copper island is shown in the 20~nm~$\times$~10~nm STM image in the inset (indicated by the blue circle). (b) 20~nm~$\times$~1~nm STM image of the copper island on bP on which the line STS measurement was performed. Spots of individual measurements are indicated by blue, green and pink dots. (c) Height profile across the island. (d)  Midgap value of individual spectra with respect to the Fermi level, plotted as a function of lateral position. The plot field is colored with yellow and blue to indicate p-type and n-type doping regions, respectively. The x-axes in (b), (c), and (d) are the same. Two vertical dashed red lines show the position of the edges of the copper island. Scan parameters of the STM images in (a) and (b) and STS setpoint for all spectra: ($-1.0$~V, $0.33$~nA). Between the spectra taken at various points along the line, the feedback loop was engaged.}
\end{figure}

To corroborate these results and determine experimentally the length scale of the electronic effect driven by copper, we performed line-spectroscopic measurements across copper islands, as shown in Figure~\ref{Fig3}. We measured in total five different copper islands, and we see line spectra showing consistent behavior. Figure~\ref{Fig3} shows a series of tunneling spectra laterally separated from each other by $\sim 2$~nm. The measurement spots follow a path across a copper island, as shown in Figure~\ref{Fig3}b. The individual spots are color-coded, from blue -- the starting point of the path, to pink -- the end point of the path, both of which lie on the flat bP surface. The copper island lies in between, with the corresponding green spots. For each spot, six spectra were recorded, and the resulting averages are shown in Figure~\ref{Fig3}a, using the same colors. The blue and pink curves, corresponding to the data from bP, are clearly different from the green curves measured on the copper island. These observations are consistent with the spectra shown in Figure~\ref{Fig2}, {thus demonstrate how reproducible these spectral features are,} and indicate a negligible drift of the tip during the line STS measurement.

A height profile across the Cu island is reported in Figure~\ref{Fig3}c, with two vertical dashed red lines marking the extremes of the island. Figures~\ref{Fig3}b-d have the same x-axis, corresponding to the scan size of the STM image. The midgap values obtained from the spectra shown in Figure~\ref{Fig3}a are plotted versus lateral position in Figure~\ref{Fig3}d. A clear transition from p-type (n-type) to n-type (p-type) behavior can be seen in correspondence of the extremes of the Cu island. The data points of Figure~\ref{Fig3}d are well fitted with a Gaussian curve, centered at $(10.78 \pm 0.17)$~nm with a FWHM of $(3.96 \pm 0.47)$~nm, in excellent agreement with the center of the copper island. Our experimental observations are in good agreement with the results shown in Figure \ref{fig:wavefuntions_7}d, and consistent with a strong quantum confinement of charge transfer from copper to bP. {Using the same technique of line-spectroscopic measurements, similar short-ranged spectral features have been previously reported for N-doped graphene \cite{Zhao2011} and P-vacancies in bP \cite{Kiraly2017}.}

Our local investigation of bP doping induced by copper islands shows that only the regions of bP lying in close proximity to copper islands are significantly n-type doped, while the regions of bP few nanometers away from the copper islands still retain their intrinsic p-type doping. Thus, the presence of isolated Cu-islands on bP implies the formation of localized n-doped regions surrounded by intrinsic bP p-doped regions. We anticipate that decreasing the copper island size and increasing their density will result in homogeneously doped n-type bP samples. Technically, this could be achieved by sputter-deposition of Cu, as already demonstrated by Koenig et al.\cite{Koenig2016} An alternative approach could be based on our result that Cu clusters preferentially nucleate at Cu$_s$ sites. Thus one might influence the distribution of Cu by a defect engineering of the P vacancies, for example by an ion bombardment of the bP surface prior to Cu deposition.  

In summary, our study provides a combined experimental and theoretical in-depth understanding of the doping behavior of Cu on bP at the nanoscale. STS measurements {suggest} Coulomb blockade in the Cu islands and an n-type doping in bP after Cu deposition, consistent with DFT predictions. The DFT analysis also shows a decoupling of the topmost bP layer in presence of copper. Line-spectroscopy measurements finally highlight that the doping effect of Cu on bP is short ranged, in good agreement with the results of our DFT calculations. This first STM investigation of Cu-doped bP  at the nanoscale is of significant importance both for a fundamental understanding of the doping mechanisms and in view of potential applications of this material in electronics. Our results indicate a route towards ultra-sharp p-n junctions in bP, which would be exciting to explore in transport measurements on high-mobility bP and might enable observation of electro-optical effects in this material.\cite{Stegmann}

\section{Methods}

{Thin flakes of bP were prepared using the scotch tape exfoliation method. The exfoliated bP flakes were transferred onto monolayer graphene grown epitaxially on silicon carbide. This graphene film  was used as a substrate and connected to ground potential. Details of sample preparation are reported elsewhere.\cite{Abhishek2018} Typical dimensions of the bP flakes are: area $(2.7 \pm 3.2)$~$\mu$m$^{2}$ and thickness $(37.5 \pm 22.4)$~nm. All measurements presented here were performed on a flake 80 nm-thick.} Copper was then deposited in-situ via thermal evaporation using an EFM-3S e-beam evaporator, and surface investigations were performed at room temperature using an Omicron LT-STM. Theoretical investigation of both stability and electronic properties of Cu-doped bP was carried out by using a first-principles plane-wave pseudopotential implementation of density functional theory (DFT), as available in the Quantum ESPRESSO package.\cite{gian+09jpcm,gian+17jpcm} The Perdew-Burke-Ernzerhof (PBE) generalized gradient approximation for the exchange-correlation functional was used,\cite{perd+96prl} and dispersion corrections were included within the semiempirical method developed by S. Grimme (DFT-D2).\cite{grim06jcc} Ultrasoft pseudopotentials were employed as available in the SSSP Library.\cite{pran+18npjcm} Further details on experiment and theory are reported in the Supporting Information.

\begin{acknowledgement}
Useful discussions with Thomas Szkopek and Herv\'e Courtois are gratefully acknowledged. FT and DP acknowledge financial support by CNR-Nano through the SEED project SURPHOS. We thank the European Research Council for funding the project PHOSFUN {\it Phosphorene functionalization: a new platform for advanced multifunctional materials} (Grant Agreement No. 670173) through an ERC Advanced Grant to MP. Computational resources were partly granted by the Center for Functional Nanomaterials at Brookhaven National Laboratory, supported by the U.S. Department of Energy, Office of Basic Energy Sciences, under Contract No. DE-SC0012704.
\end{acknowledgement}

\begin{suppinfo}
The following files are available free of charge.
\begin{itemize}
  \item Supporting Information: Extended experimental methods, Raman spectroscopy measurements, STM images of bP surfaces before and after Cu deposition, Stability of Cu islands, Numerical details of DFT simulations and complementary data.
\end{itemize}
\end{suppinfo}

\bibliography{ref}

\begin{mcitethebibliography}{64}
\bibitem[Not()]{Note-2}
We note that structural optimization gives rise to a sliding of the layers in
  presence of interstitial Cu and a subsequent gap opening due to a partial
  decoupling of the layers. This behaviour is not expected to show up in real
  samples. We have thus chosen to mimic real samples by inhibiting layer
  sliding. In the latter case we find no significant gap opening\relax
\mciteBstWouldAddEndPuncttrue
\mciteSetBstMidEndSepPunct{\mcitedefaultmidpunct}
{\mcitedefaultendpunct}{\mcitedefaultseppunct}\relax
\EndOfBibitem
\end{mcitethebibliography}

\end{document}